\documentclass[lettersize,journal]{IEEEtran}
\usepackage{amsmath,amsfonts}
\usepackage{algorithmic}
\usepackage{algorithm}
\usepackage{array}
\usepackage[caption=false,font=normalsize,labelfont=sf,textfont=sf]{subfig}
\usepackage{textcomp}
\usepackage{stfloats}
\usepackage{url}
\usepackage{verbatim}
\usepackage{graphicx}
\usepackage{cite}
\usepackage{bbding}
\usepackage{hyperref}
\usepackage{amssymb}
\begin{document}

\title{A noise-tolerant, resource-saving probabilistic binary neural network implemented by the SOT-MRAM compute-in-memory system
}

\author{Yu Gu, Puyang Huang, Tianhao Chen, Chenyi Fu, Aitian Chen, Shouzhong Peng, Xixiang Zhang and Xufeng Kou,~\IEEEmembership{Senior Member,~IEEE,}

\thanks{This work is supported by the National Key R\&D Program of China (2021YFA0715503), the NSFC Programs (92164104), and the Shanghai Rising-Star Program (21QA1406000). \textit{(Corresponding author: Xufeng Kou)}}

\thanks{Gu Yu, Puyang Huang, and Xufeng Kou are with the School of Information Science and Technology, ShanghaiTech University, Shanghai 201210, China (e-mail: kouxf@shanghaitech.edu.cn).}
\thanks{Tianhao Chen, Chenyi Fu and Shouzhong Peng are with the with the School of Integrated Circuit Science and Engineering, Beihang University, Beijing 100191, China.}
\thanks{Aitian Chen and Xixiang Zhang are with the with Physical Science and Engineering Division, King Abdullah University of Science and Technology, Thuwal, 23955-6900, Saudi Arabia.}
}

\maketitle

\begin{abstract}
We report a spin-orbit torque(SOT) magnetoresistive random-access memory(MRAM)-based probabilistic binary neural network(PBNN) for resource-saving and hardware noise-tolerant computing applications. With the presence of thermal fluctuation, the non-destructive SOT-driven magnetization switching characteristics lead to a random weight matrix with controllable probability distribution. In the meanwhile, the proposed CIM architecture allows for the concurrent execution of the probabilistic vector-matrix multiplication (PVMM) and binarization. Furthermore, leveraging the effectiveness of random binary cells to propagate multi-bit probabilistic information, our SOT-MRAM-based PBNN system achieves a 97.78\% classification accuracy under a 7.01\% weight variation on the MNIST database through 10 sampling cycles, and the number of bit-level computation operations is reduced by a factor of 6.9 compared to that of the full-precision LeNet-5 network. Our work provides a compelling framework for the design of reliable neural networks tailored to the applications with low power consumption and limited computational resources.
\end{abstract}


\section{Introduction}
\IEEEPARstart{A}{rtifacial} Neural Networks (ANNs) have become an indispensable driving force to orchestrate significant advancements in image recognition, visual intelligence, and natural language processing. Unfortunately, the integration of conventional full-precision ANNs into edge and mobile systems remains as a critical challenge as they invariably demand substantial computational resources\cite{ref1}. In response to the ever-growing computationally intensive tasks, the concept of binary neural networks (BNNs), which utilizes binary weights to construct lightweight networks, has been proposed to accelerate the data processing, yet at the expense of degraded performance\cite{ref2}. Inherited from the resource-saving nature of BNN, the probabilitic binary neural networks (PBNNs) use random binary variables as weight information to retain more input details so that high classification accuracies can be realized with limited sampling cycles\cite{ref3}. Additionally, the introduced randomness during the training and inference processes may endow PBNNs with a salient noise-tolerant feature, rendering them suitable for tasks involving significant noise and contaminated datasets.

Generally, the key to construct a reliable PBNN lies in the generation of a set of random binary bits with a controllable distribution pattern. In this context, software-based pseudo-random number generators often lag in stochastic algorithm performance\cite{ref5}. On the other hand, while CMOS hardwares can generate true random bits by leveraging the thermal noise, such approaches necessitate complicated circuitries with large layout areas, considerable power consumptions and additional digital signal processing to adjust the distribution of the random outputs\cite{ref6}\cite{ref7}. Alternatively, thanks to the thermal fluctuation induced magnetic domain motion, magnetoresistive random-access memory (MRAM) device is an ideal candidate for true random number generators (TRNGs)\cite{ref8}. More importantly, benefiting from the spin precession dynamics triggered by spin-orbit torque (SOT), the switching probability of SOT-MRAM can be precisely controlled by the programmable input current\cite{ref9}. In addition, the separated read and write paths may, in principle, enable the three-terminal SOT-MRAM bit-cell as a compelling building block for implementing an energy-efficient PBNN system with a computing-in-memory (CIM) architecture\cite{ref11}.

\begin{figure}[!t]
    \centering
    \includegraphics[width=3.5in]{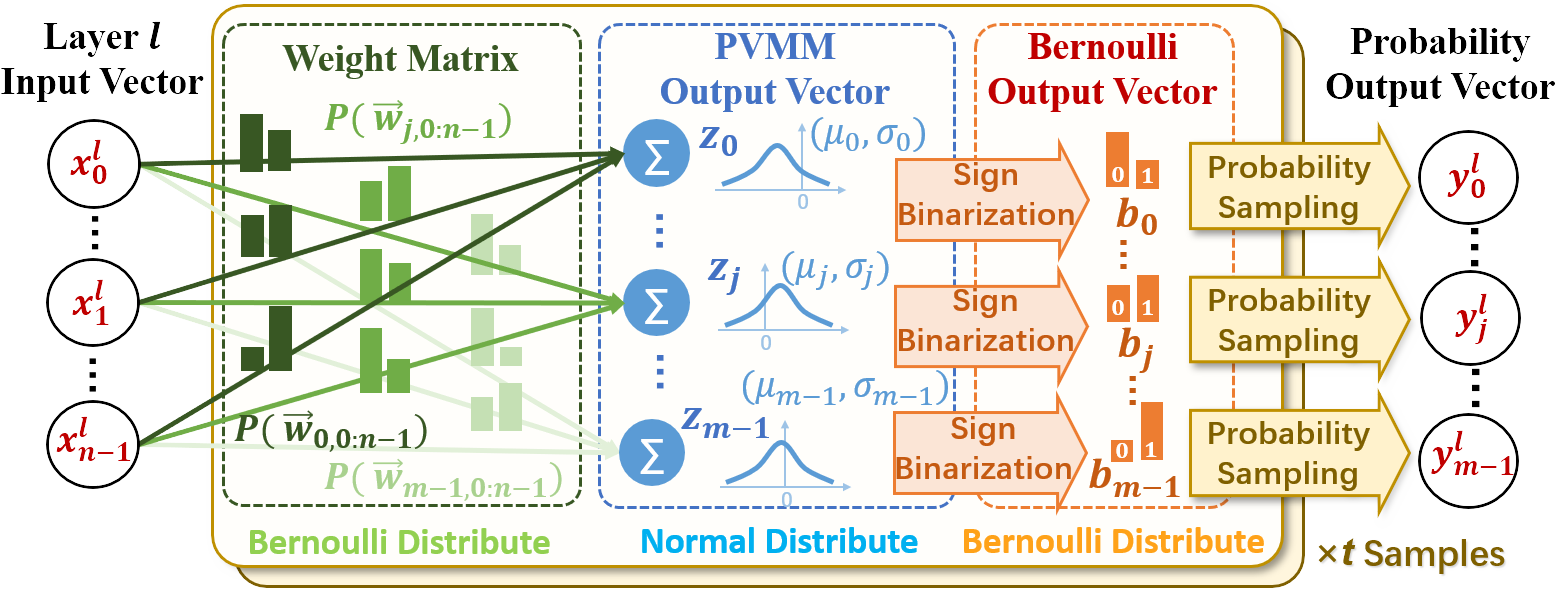}
    \caption{Schematic of the layer in probabilistic binary neuron networks.}
    \label{fig_1}
\end{figure}

Inspired by the aforementioned merits, in this brief, we report the SOT-MRAM based PBNN system for resource-saving and noise-tolerant MNIST classification. The remainder of this brief is organized as follows: Sections II, III, and IV delve into the algorithm, circuit design, and performance evaluation of the MRAM CIM chip, respectively. The concluding Section V summarizes the key findings.

\section{Theory of Ideal Probabilistic Binary Neural Network}
\subsection{Probabilistic Vector-Matrix Multiplication Algorithm}
Unlike traditional deterministic neural networks, the PBNN encodes random binary bits as the probability elements $w_{ij}$ (i.e., which follows the Bernoulli distribution $Bern(w_{ij})$) in the random weight matrix $[W_{0:m-1,0:n-1}]$, as illustrated in Fig. 1. On this basis, the element in the probabilistic vector-matrix multiplication (PVMM) vector $\overrightarrow{z}_{0:m-1}$ (i.e., the dot product of the random weight matrix and the static input vector $\overrightarrow{x}_{0:n-1}$) obeys the Normal distribution $\mathcal{N}(\mu_{j},\sigma_{j}^2)$ according to the (Lyapunov) Central Limit Theorem (CLT):
\begin{equation}
    z_j = \overrightarrow{x}_{0:n-1} \cdot \overrightarrow{W}_{0:n-1,j} \sim \mathcal{N}(\mu_{j},\sigma_{j}^2) \\
\end{equation}
$$ 
\begin{cases}
\mu_{j} & = \frac{1}{n}\sum_{i=0}^{n-1}(2w_{ij}-1)\cdot x_i\\
\sigma_{j}^2 & = \frac{1}{n-1}\sum_{i=0}^{n-1}4w_{ij}(w_{ij}-1)\cdot x_i^2
\end{cases}$$
where $\mu_j$ and $\sigma_j^2$ are the mean and variance of the Normal distribution, respectively. Afterwards, the binary output vector $\overrightarrow{b}_{0:m-1}$ is obtained by extracting the sign bits of $\overrightarrow{z}_{0:m-1}$ (i.e., sign binarization). Accordingly, the expectation of the element $b_j$ (i.e., which also obeys Bernoulli distribution relation) represents the likelihood of $z_j > 0$, and the corresponding probability output vector $\overrightarrow{y}_{0:m-1}$, is finally generated through the successive sampling of such binarized $\overrightarrow{b}_{0:m-1}$ result. Based on this procedure, the PBNN manages to preserve more input details within the binary weight cells, thus achieving good performance with limited hardware resource.

\begin{figure}[!t]
\centering
\includegraphics[width=3.2in]{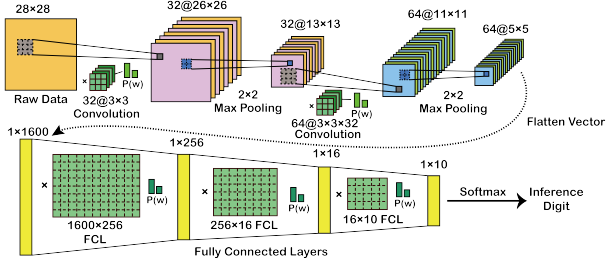}
\caption{Proposed PBNN framework for the MNIST test where the network consists of 2$\times$ Convolution Layers (Convs), 2$\times$ Max Pooling Layers (MPs), 3$\times$ Fully Connected Layers (FCLs) and the Softmax Layer.}
\label{fig_2}
\end{figure}

\begin{figure}[!t]
\centering
\includegraphics[width=3.5in]{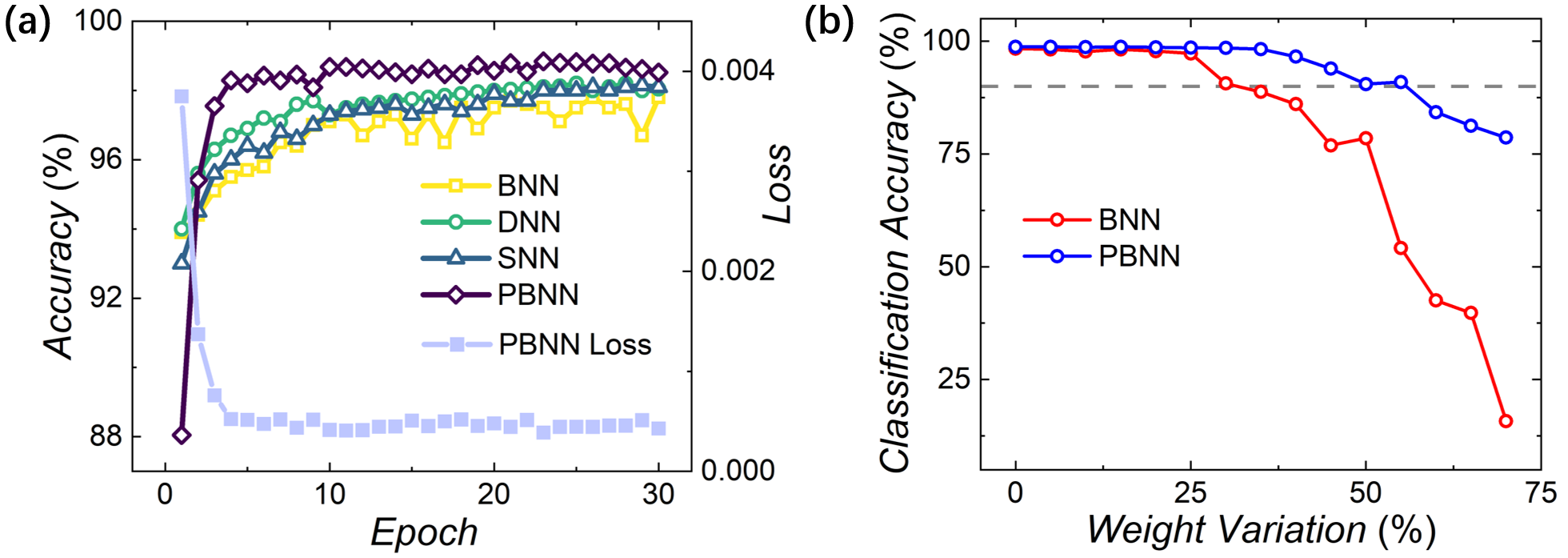}
\caption{(a). MNIST training results versus epoch times among different networks (b). Comparison of the classification accuracy between BNN and PBNN against weight variation.}
\label{fig_3}
\end{figure}

\subsection{Training result of the ideal PBNN system}
To evaluate the performance of PBNN, we designed a standard convolution neural network (CNN) based on the proposed PBNN algorithm and carried out the handwritten digit classification task using the MNIST dataset. As specified in Fig. 2, our PBNN network consists of two sets of Convolution and Max-pooling layers and three Fully Connected layers. Except for the first layer, all the input and output data in this network are in the form of binary bits. Based on this framework, our PBNN (432 Kb parameters) involves 5.7 million bit multiplication and adding operations during each sampling cycle,(i.e., the total sampling cycle was set at 10). Recall that the full-precision CNN network of LeNet-5 structure contains 2.016 Mb parameters and 395 million bit operations\cite{cnnref1} , our system thereafter manages to reduce the number of bit-level calculations by a factor of 6.9.

\begin{figure}[!t]
\centering
\includegraphics[width=3.5in]{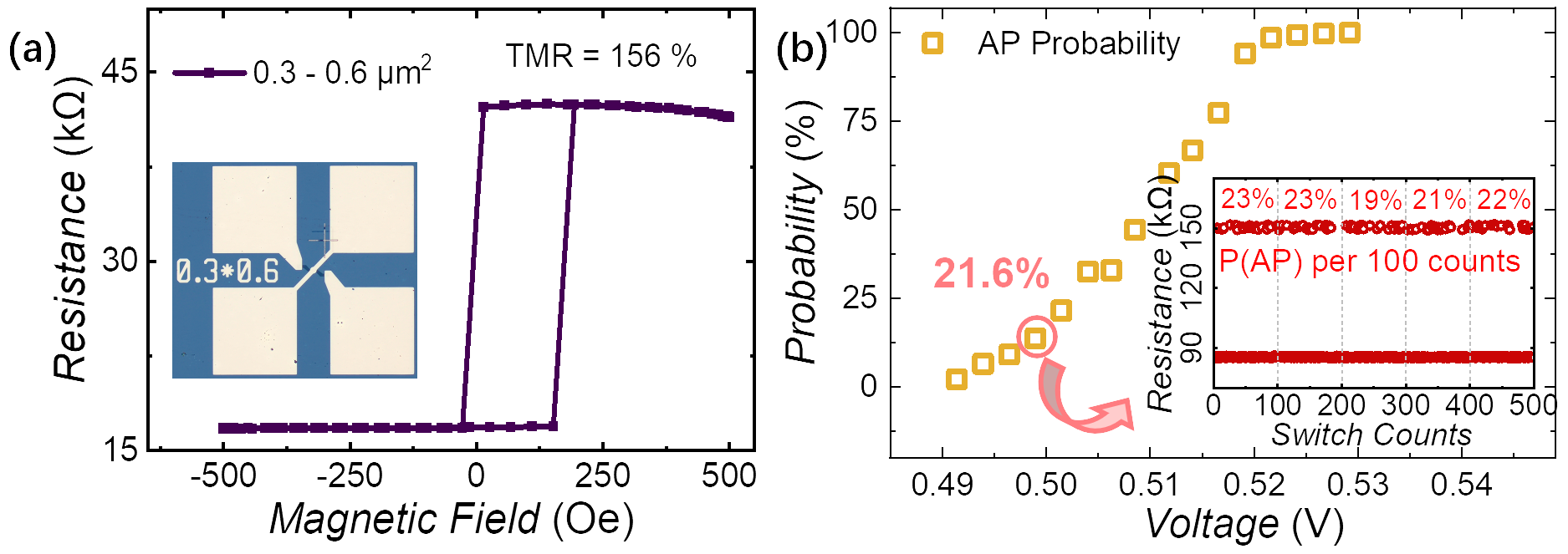}
\caption{(a). The tunnel magneto-resistance of the in-plane SOT-MRAM device at room temperature. (b). Measured switching probability curve of the SOT-MRAM cell versus the write voltage ($V_{wr}$). The statistical probability at each point was obtained after 500 sampling cycles.}
\label{fig_5}
\end{figure}

Accordingly, quantitive comparisons of the simulation results on various neural networks are summarized in Fig. 3a \cite{nnref1}\cite{nnref2}\cite{nnref3}. It is seen that the classification accuracy of the ideal PBNN network (i.e., which assumes the distortionless sampling of probabilities with infinite sampling cycles) quickly exceeds 98\% within just 5 epochs, and it converges to the 98.4\% baseline after 20 epochs, manisfesting its fast-training trait compared to other nueral networks. Strikingly, the training result of our PBNN system remains almost constant against the weight variation lower than 25\% (i.e., which is primarily attributed to circuit noise and the device-to-device variation of memory cells, and hence plays an important role in the analog accumulation in CIM system), and Fig. 3b unveils that the classification accuracy is still above 90\% even when the weight variation is up to 50\% (i.e., in contrast, the accuracy of the deterministic BNN counterpart drops to 75\% at the same weight variation level). In conclusion, these theoretical simulation results validate the advantage of our proposed PBNN in terms of high training speed and good noise-tolerant feature.

\begin{figure*}[!t]
    \centering
    \includegraphics[width=7in]{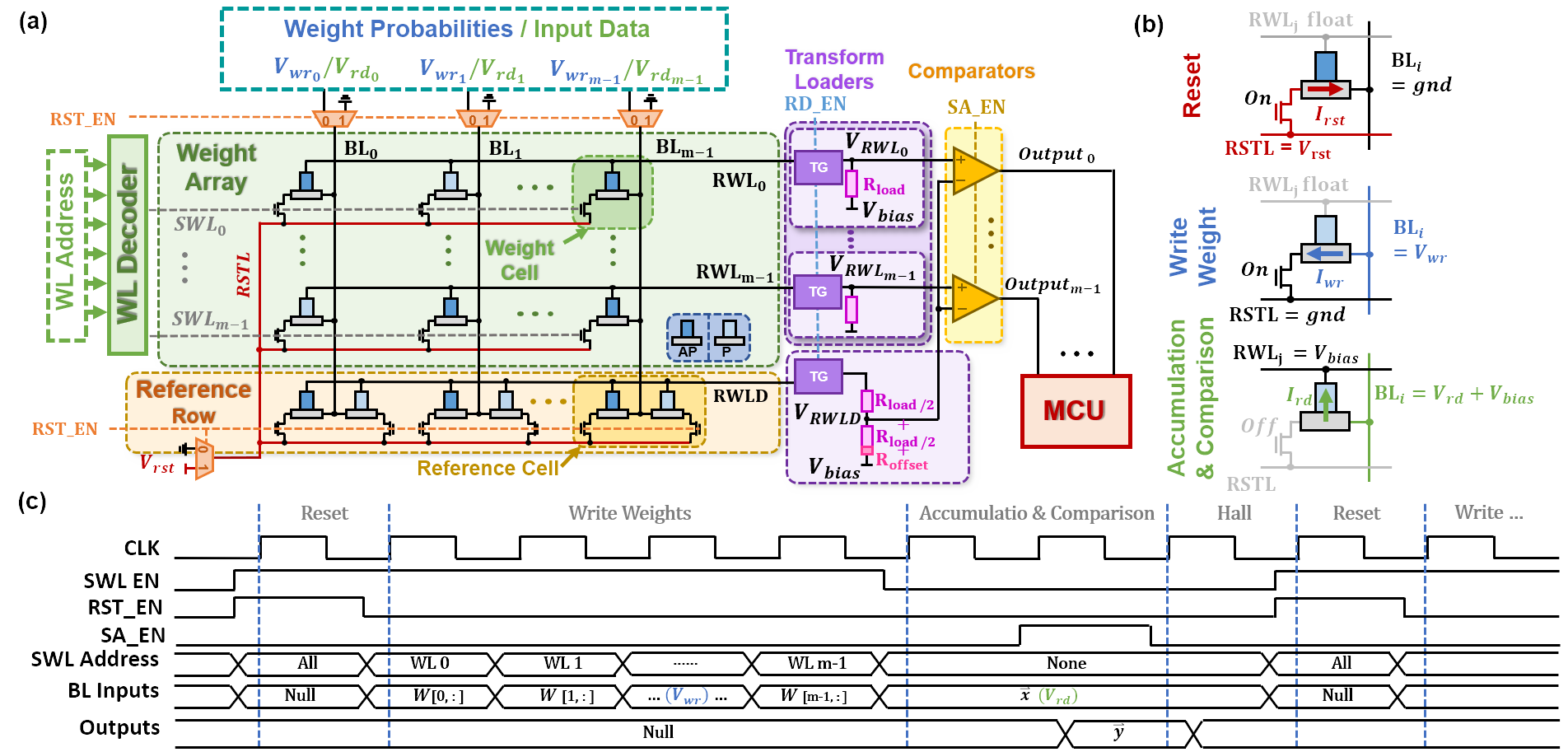}
    \caption{Hardware Implementation of the SOT-MRAM-based PBNN system. (a). Circuit Schematic of the SOT-MRAM CIM module. (b). Input voltage patterns in reference to different operation modes of the bit-cell. (c). Operation waveforms of the PBNN CIM chip. }
    \label{fig_6}
    \end{figure*}

\section{Hardware Implementation of PBNN}
\subsection{SOT-MRAM device characterization}
As the basic element of the PBNN, the three-terminal SOT-MRAM device uses the spin current generated in the bottom heavy metal layer to manipulate the magnetization of the adjacent magnetic free layer via the spin-orbit torque, which in turn changes the tunnel magneto-resistance (TMR) of the top magnetic tunnel junction (MTJ) structure \cite{ref13}. In this regard, facilitated by this non-destructive SOT-driven magnetization switching mechanism, the repeatable switching probability of the SOT-MRAM device can be well-controlled by the input signal under identical thermal noise conditions. Fig. 4a exemplifies the device characteristics of one typical SOT-MRAM device with the MTJ size of 300 nm $\times$ 600 nm. The room-temperature TMR ratio is calibrated as 156\%, which warrants a sufficient readout margin between the anti-parallel (AP) and parallel (P) states for the following PVMM accumulation process. Meanwhile, when the device is stimulated by a series of input signals $V_{wr}$ with the fixed pulse width of 40 ms, the corresponding switching probability curve is disclosed in Fig. 4b. For instance, after 500 times sampling, it is observed that the resistance state of the SOT-MRAM bit-cell has flipped 108 times under a given $V_{wr} = 0.5 V$, which equals to a switching probability of 21.6\% (inset of Fig. 4b). More importantly, such a switching probability value produced by the same $V_{wr}$ is found to be highly consistent in 5 sub-groups (i.e., each of which includes 100 sampling cycles), with the standard deviation of 1.67\%, hence justifying the repeatability of the SOT-MRAM device. Accordingly, the aforementioned results were incorporated into a device compact model coded in Veriog-A (i.e., other key parameters are listed in Table I), which was subsequently applied on the following circuit-level simulations using the 40nm CMOS technology process design kits.

\begin{table}[!h]
    \caption{Key Device Parameters of SOT-Device Adopted for Simulation\label{tab:table_2}}
    \centering
    \resizebox{\linewidth}{!}{
    \begin{tabular}{|c|c|c|c|c|c|}
    \hline
    \bf{$RA $} & Resistance area product & $\Omega\cdot m^2 $ & 1.51 $\times 10^{-9} $ \\
    \hline
    \bf{$t_{sl} $} & Free layer thickness & nm & 2.5 \\
    \hline
    \bf{$t{ox} $} & MgO barrier thickness & nm & 1.7 \\
    \hline
    \bf{$V_h $} & Voltage bias when TMR(V) = 0.5$\times $TMR(0) & V & 0.75 \\
    \hline
    \bf{$TMR $} & TMR ratio under zero bias voltage & \% & 156 \\
    \hline
    \bf{$a $} & MTJ pillar length & nm & 600 \\
    \hline
    \bf{$b $} & MTJ pillar width & nm & 300 \\
    \hline
    \bf{$d $} & HM-strip thickness & nm & 6 \\
    \hline
    \bf{$\rho $} & HM-strip resistivity & $\Omega\cdot $m & 195 $\times 10^{-8} $ \\
    \hline
    \end{tabular}
    }
\end{table}

\subsection{SOT-MRAM based CIM Circuit Design}
Next, the  SOT-MRAM-baseed CIM chip  was designed to enable the energy-efficient PBNN operation. As displayed in Fig. 5a, this CIM structure includes the weight array, the reference row, the WL decoder, the transform loaders, and the comparators. According to the PBNN flow diagram (Fig.1), the weight matrix sampling and PVMM are implemented in the weight array, and Fig. 5b-c show the bit-cell bias conditions and relevant waveforms in reference to the Reset, Write Weight, Accumulation \& Comparison operations, respectively. In particular, after all SOT-MRAM devices are initialized at the high-resistive AP state in the Reset cycle, the $SWL_j$ and $BL_{0:m-1}$ signals trigger the Write operation on the selected row of SOT-MRAM cells, where the input $V_{wr}$ would inject the resulting charge current through the bottom heavy-metal layer and tailor the MTJ magnetization configuration (i.e., either stays at AP or switches to the P state) with respect to the switching probability curve of Fig. 4b. By successively repeating the above Reset and Write cycles with the same input patterns, the weight sampling operation is accomplished.

In view of the Accumulation \& Comparison cycles, the readout voltage ($V_{rd}$) is firstly applied on the MTJ structure to extract the weight information (i.e., the conductance of the MTJ $G_{i,j}$). In the meanwhile, all the transistor drivers are turned off by $SWL_j$ to avoid accidental writing. Based on the Kirchhoff's Current Law (KCL) mechanism, the output currents of the parallel connected SOT-MRAM bit-cells along the same row are directly summed up ($I_{RWL_j}=\sum_i G_{i,j} \times V_{rd_i}$) so that the PVMM operation is automatically achieved. This PVMM current is subsequently converted to the PVMM voltage $V_{RWL_j}$ in the Transform Loader unit, which corresponds to the normal distributed PVMM output $z_j$ in the PBNN algorithm (Fig. 1), and the sign bit of $z_j$ is determined by comparing $V_{RWL_j}$ with the reference value $V_{RWLD}$ generated in the reference row module. Moreover, to attain a proper binary partition of the PVMM results under different input cases, we have introduced a reference cell (which consist of a pair of SOT-MRAM devices with opposite P and AP states, so that the reference point is kept at the average value). In the last step, the auxiliary MCU will count and average the outputs of the comparators, hence obtaining the final probability outputs in reference to different sampling cycles.

\begin{figure}[!t]
    \centering
    \includegraphics[width=3.2in]{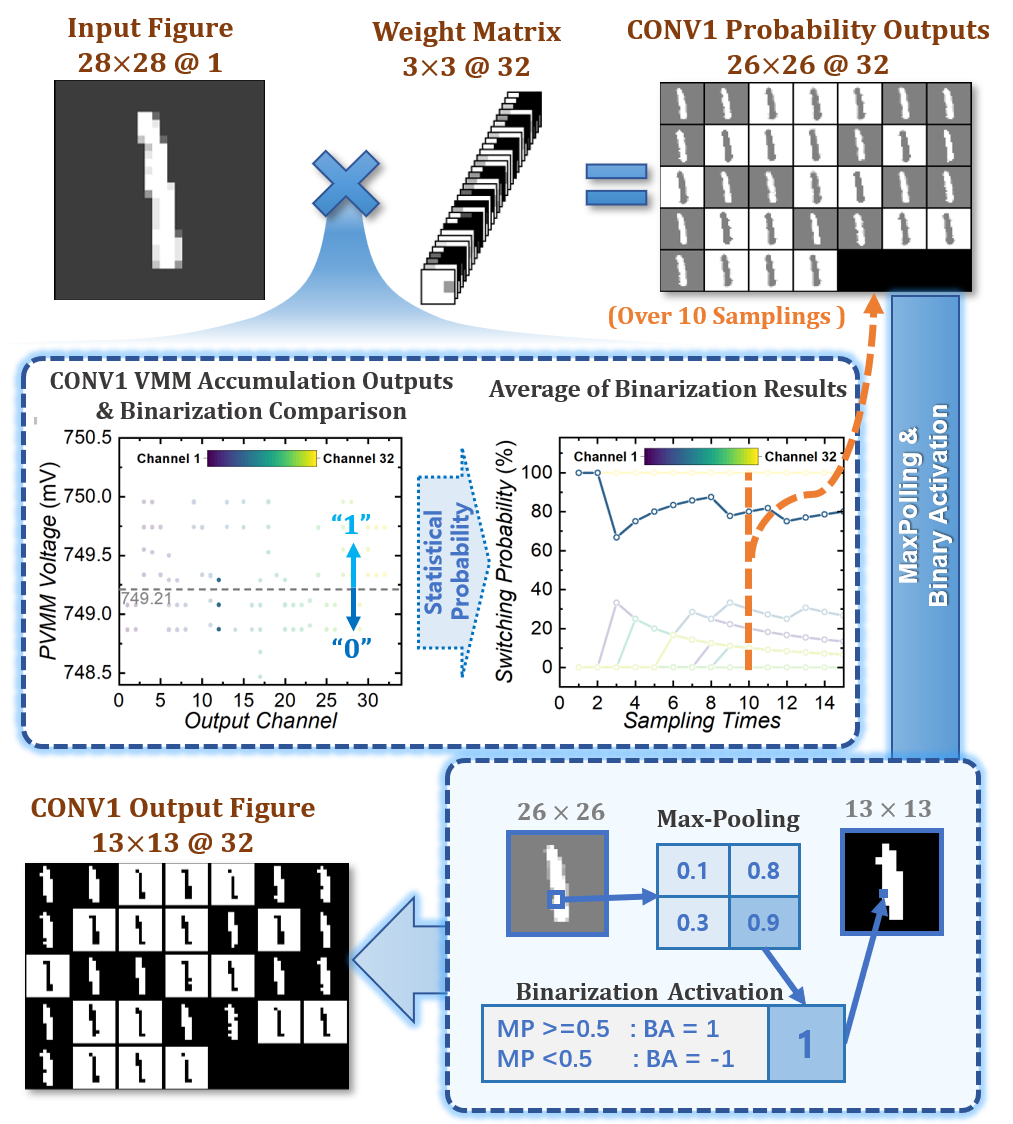}%
    \caption{Circuit simulation results of the first convolution layer (CONV-1), in the PBNN network in response to the input figure of digit "1".}
    \label{fig_7}
\end{figure}

\begin{figure}[!t]
    \centering
    \includegraphics[width=3.2in]{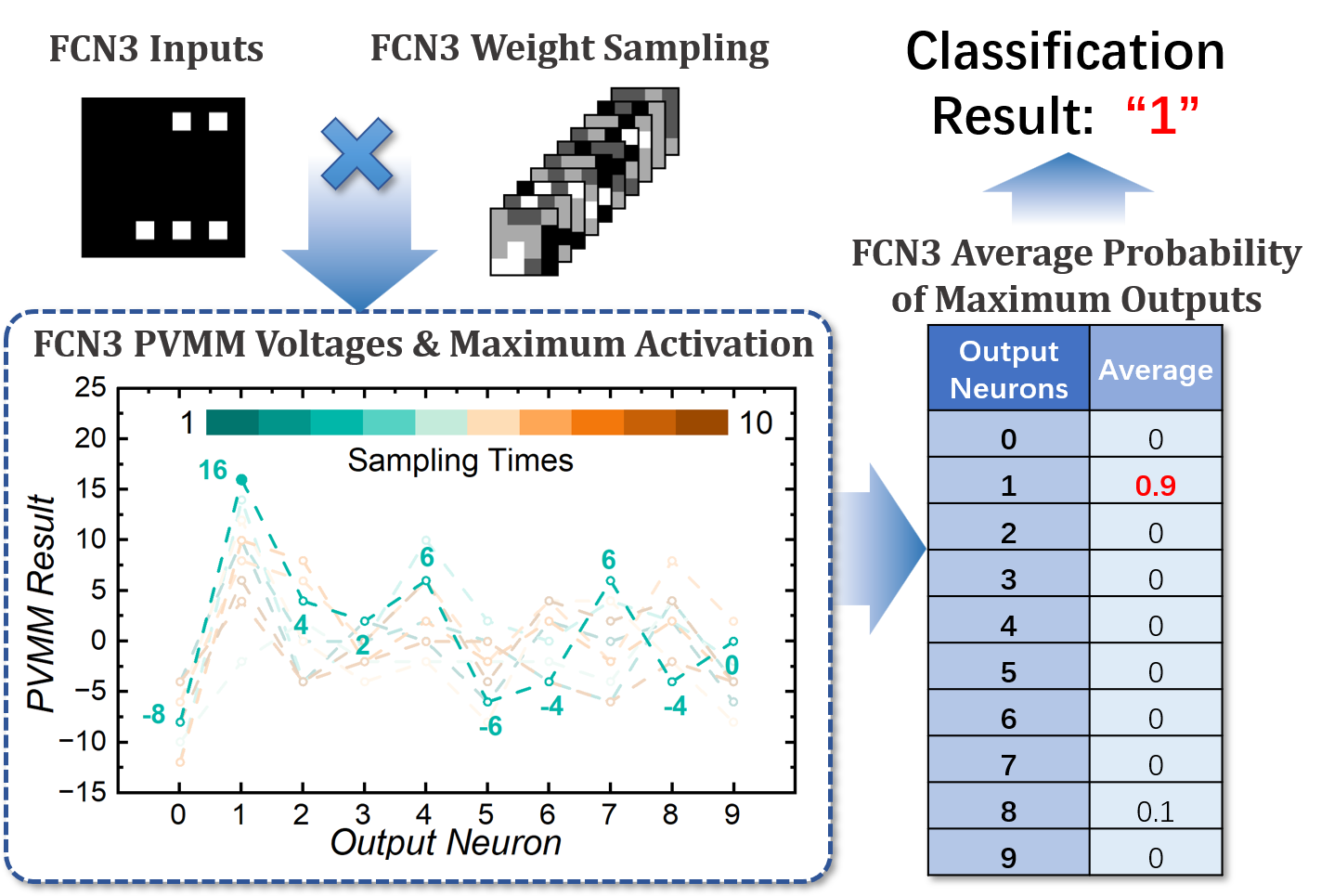}%
    \caption{ The average Maximum Activation expectations of 10 output neurons after 10 sampling cycles and the final handwritten digit classification result of the SOT-MRAM-based PBNN system.}
    \label{fig_8}
    \end{figure}

\section{Simulation Results and Analysis of the SOT-MRAM-based PBNN Hardware}
Following the architecture elaborated above, we have constructed a 32$\times$9 SOT-MRAM array as the first layer of the PBNN system. The input case of inferring digit "1" and the weight kernels in CONV1 layer are visualized in Fig. 6, where the inputs and weights both have 4 quantization states for the circuit-level simulation.

Specifically, the PVMM results of different channels are represented by the PVMM voltage $V_{RWL_j}$ ($j = 0$ to 31), which is updated along with the weight matrix after each sampling cycle. As exemplified in the middle panel of Fig. 6, with the presence of thermal fluctuation (i.e., which introduces stochasticity in the weight matrix), the recorded $V_{RWL_i}$ values are found to range from 748.5 mV to 750 mV. Given that the $V_{RWLD}$ value of the reference cell is 749.2 mV (dotted line), the sign of the PVMM result from each channel can be therefore labeled as "1" or "0". Under such circumstances, the mean value of this binarization output gradually converges to a constant after 10 sampling cycles. Therefore, in order to make an appropriate balance between the accuracy and power consumption, we adopt this value as the binarization probability for the following process. Accordingly, the gray image of probabilities (26$\times$26@32) propagated through the max-pooling layer (i.e., whose size of the kernel is 2$\times$2), and were futher scaled down to 13$\times$13@32. Afterwards, the binarization activation was applied on the output image of the max-pooling layer, which in turn yielded the final binary outputs of this layer, as shown in the bottom panel of Fig. 6 (i.e., where the black pixel in the CONV-1 Output Figure represents "-1" and the white one represents "1").

\begin{figure}[!t]
    \centering
    \includegraphics[width=3.5in]{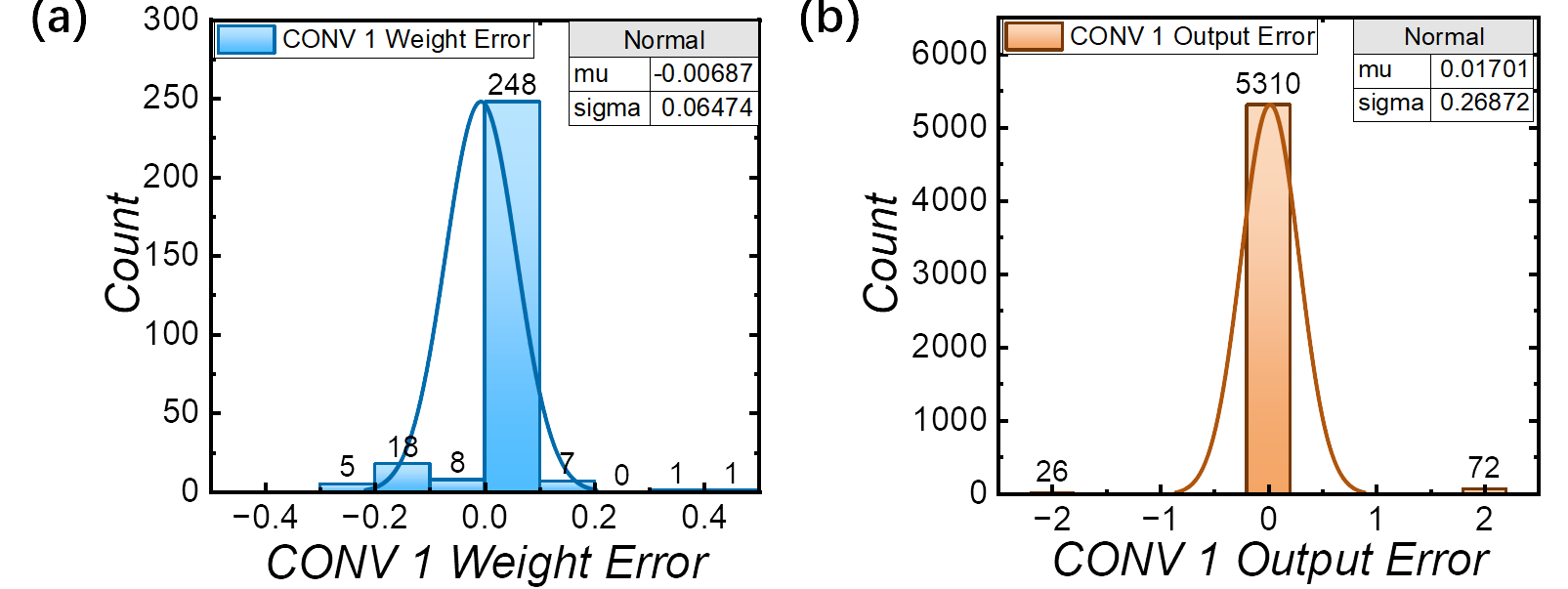}%
    \caption{Error distribution of (a) weight probabilities and (b) binary outputs of the CONV 1 layer with 10 sampling cycles in the SOT-MRAM-based PBNN system compared to the ideal baseline of PBNN with 4-state quantization of inputs and weights.}
    \label{fig_9}
    \end{figure}

Following the same procedure, the data procession on other convolution (CONV) and fully-connected (FCN) layers was simulated by Matlab, and Fig. 7 displays the handwritten digit classification result from the last FCN layer with 16 binary inputs. Taking one sampling cycle as an example, the neuron at $2^{nd}$ channel has the largest PVMM voltage of 10, thus the Maximum Activation output of this nueron is one and the outputs of other nuerons are zero in this sampling cycle. After 10 samples, the average Maximum Activation expectation at the $2^{nd}$ channel is 0.9 (i.e., the maximum PVMM voltage appears in this channel for 9 times out of the total 10 samplings),which indicates that the most likely inference result is digit 1, therefore obtaining the correct recognition from the MNIST database.

In addition to the validated operating principle, the disparity between the simulation results of the SOT-MRAM-based PBNN hardware and the ideal scenario has also been evaluated. Fig. 8a presents the error distribution spectrum between the weight probabilities averaged by 10 sampling cycles and the ideal values (i.e., which are sampled by infinite cycles) with 4-state quantization, where more than 97\% of weight errors are negligible ($<$ 0.2) according to the weight variation assessment in Fig. 3b. Concurrently, the distribution of the binary output results in the CONV-1 layer is also highly consistent with the ideal curve with a negligible error rate of 1.68\% (i.e., 98 error bits out of 5408 output bits, as shown in Fig. 8b). Furthermore, we have assessed the sampling loss of the whole PBNN network compared to the ideal case with 4-state quantization, and Fig. 9 summarizes the mean square errors (MSEs) of the weight probability and output error rate of each constituent layer. It is seen that the calibrated MSE values and output error rates in all layers are smaller than 10\% as illustrated in Fig. 3b. Here, it is worth noting that even though the FCN-1 layer has a relatively high error rate of 8.79\%, the noise-tolerant feature of PBNN can effectively mitigate the loss as the output size dramatically decreases in the fully connected layer (i.e., high-level information extraction) and the loss eventually diminishes to 0\% in the FCN-2 layer. Nevertheless, we need to point out that the presence of excessive output noise may propagate through adjacent layers and disrupt the information extraction process, thereby introducing a small degree of uncertainty to the final training results.
    
\begin{figure}[!t]
    \centering
    \includegraphics[width=3.5in]{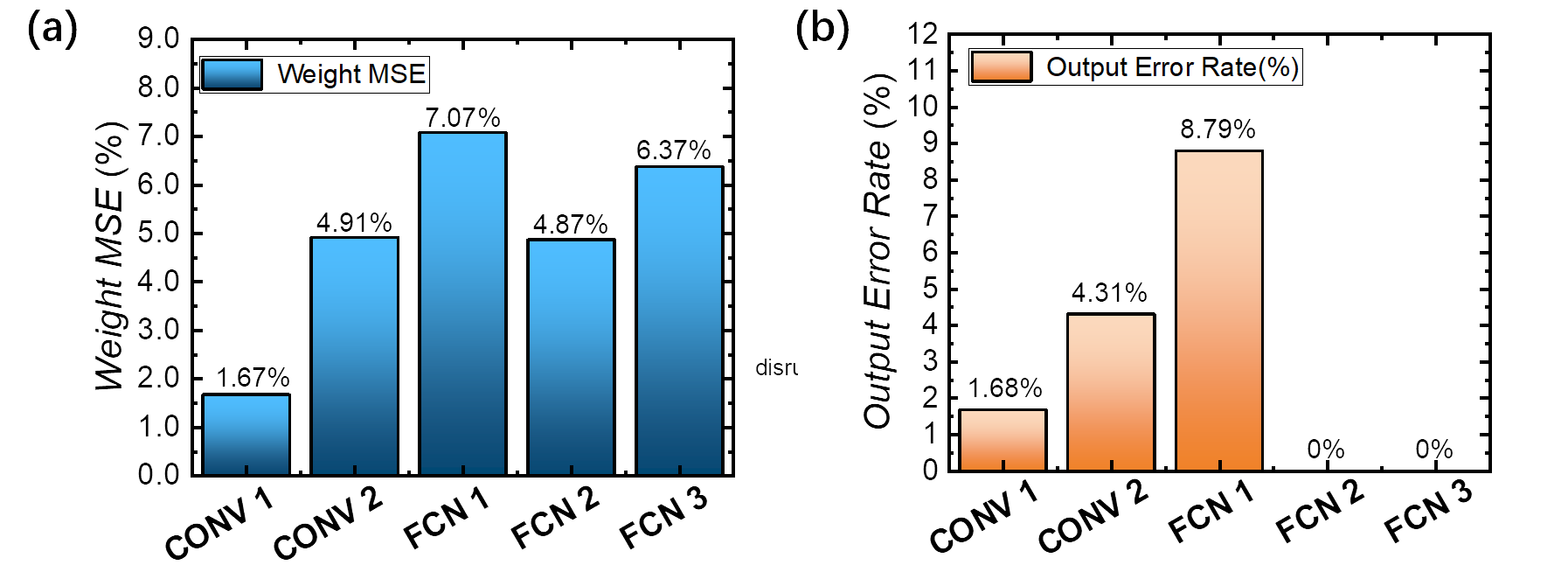}%
    \caption{Mean squared errors of the (a) weight probabilities and (b) output error rates at different layers with 10 sampling cycles in recognizing the handwritten digit 1.}
    \label{fig_10}
\end{figure}

Based on these statistical analysis, we further performed the MNIST data training on the SOT-MRAM-based PBNN, and investigated the classification accuracy as a function of the sampling cycle. As shown in Fig. 10, our system manages to achieve a 96.94\% accuracy after only 5 sampling cycles. With the increase in the cycle number (i.e., which results in a reduction in the variance of the sampled weight probabilities), the classification accuracy progressively improves, and it finally approaches the ideal baseline of 98.4\%, albeit with the cost of proportionally increased execution time and power consumption. In this context, we used the product of the accuracy error and sampling cycle as the benchmark to evaluate the overall inference performance. As depicted in Fig. 10, the product reaches the minimum value when the sampling cycle is 10 (i.e., which corresponds to the weight probability variation of 7.01\% across the entire network), hence highlighting the optimal point for achieving the energy-efficient and high-precision PBNN training.

\section{Conclusion}
In conclusion, we have designed a SOT-MRAM based lightweight PBNN system for MNIST test. Simulation results from PyTorch indicate that our PBNN model exhibits an improved noise tolerance compared to traditional BNNs. Furthermore,the SOT-driven switching dynamics gives rise to the controllable switching probability curves which help to construct the high-quality weight matrix with a low write error rate. Therefore, our results underscore the advantages of the PBNN hardware for achieving enhanced classification accuracy with fewer computation resources, hence providing a promising strategy for low-power applications. 

\begin{figure}[!t]
    \centering
    \includegraphics[width=3.5in]{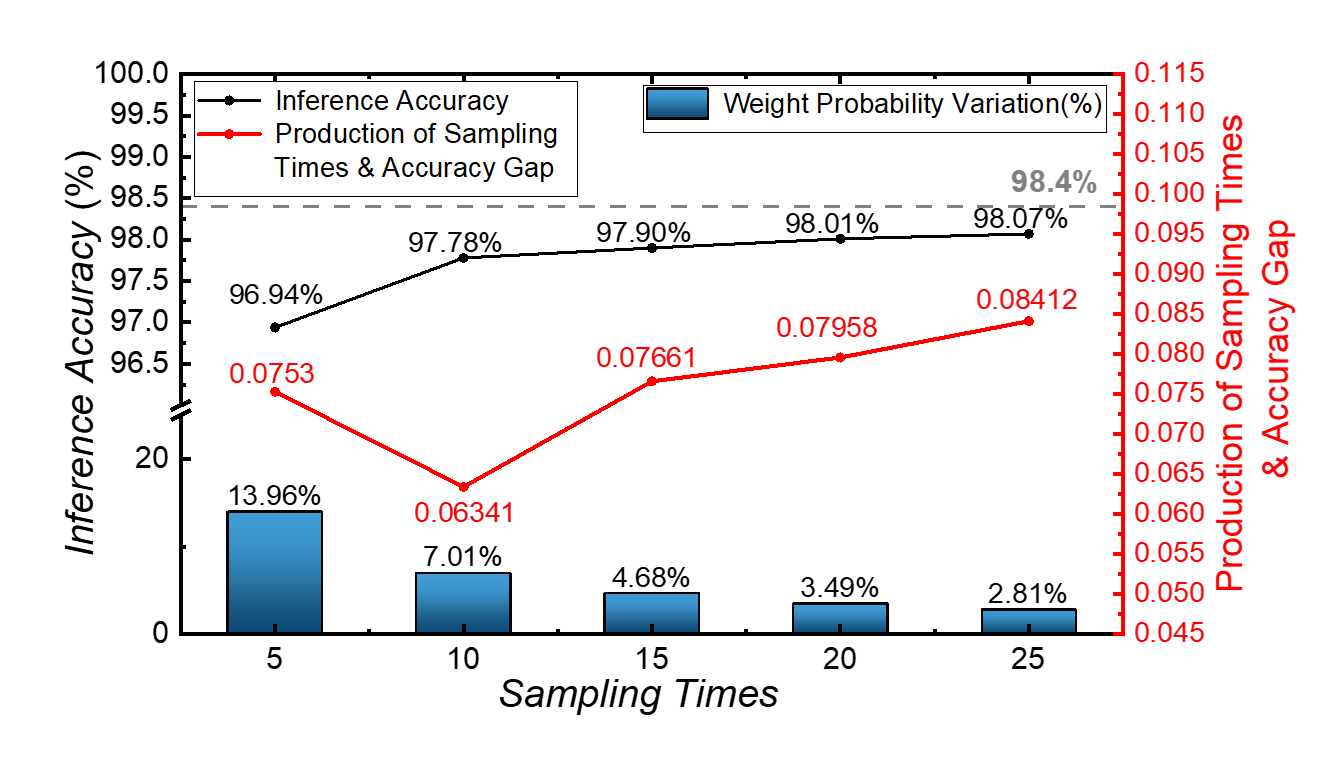}%
    \caption{The network inference accuracy with various sampling cycles and the corresponding weight variation. }
    \label{fig_11}
    \end{figure}

\bibliography{ref_0.bib}

\begin{thebibliography}{10}
\providecommand{\url}[1]{#1}
\csname url@samestyle\endcsname
\providecommand{\newblock}{\relax}
\providecommand{\bibinfo}[2]{#2}
\providecommand{\BIBentrySTDinterwordspacing}{\spaceskip=0pt\relax}
\providecommand{\BIBentryALTinterwordstretchfactor}{4}
\providecommand{\BIBentryALTinterwordspacing}{\spaceskip=\fontdimen2\font plus
\BIBentryALTinterwordstretchfactor\fontdimen3\font minus \fontdimen4\font\relax}
\providecommand{\BIBforeignlanguage}[2]{{%
\expandafter\ifx\csname l@#1\endcsname\relax
\typeout{** WARNING: IEEEtran.bst: No hyphenation pattern has been}%
\typeout{** loaded for the language `#1'. Using the pattern for}%
\typeout{** the default language instead.}%
\else
\language=\csname l@#1\endcsname
\fi
#2}}
\providecommand{\BIBdecl}{\relax}
\BIBdecl

\bibitem{ref1}
S.~W. Keckler, W.~J. Dally, B.~Khailany, M.~Garland, and D.~Glasco, ``Gpus and the future of parallel computing,'' pp. 7--17, 2011.

\bibitem{ref2}
Y.~Fujiwara and T.~Kawahara, ``Bnn training algorithm with ternary gradients and bnn based on mram array,'' in \emph{TENCON 2023 - 2023 IEEE Region 10 Conference (TENCON)}, 2023, pp. 311--316.

\bibitem{ref3}
\BIBentryALTinterwordspacing
J.~W.~T. Peters and M.~Welling, ``Probabilistic binary neural networks,'' \emph{ArXiv}, vol. abs/1809.03368, 2018. [Online]. Available: \url{https://api.semanticscholar.org/CorpusID:52180104}
\BIBentrySTDinterwordspacing

\bibitem{ref5}
F.~Neugebauer, I.~Polian, and J.~P. Hayes, ``Building a better random number generator for stochastic computing,'' in \emph{2017 Euromicro Conference on Digital System Design (DSD)}, 2017, pp. 1--8.

\bibitem{ref6}
P.~Monteiro, L.~Oliveira, and J.~Casaleiro, ``True random number generator implemented in 130 nm cmos nanotechnology,'' in \emph{2022 International Young Engineers Forum (YEF-ECE)}, 2022, pp. 52--56.

\bibitem{ref7}
T.~Arciuolo and K.~M. Elleithy, ``Parallel, true random number generator (p-trng): Using parallelism for fast true random number generation in hardware,'' in \emph{2021 IEEE 11th Annual Computing and Communication Workshop and Conference (CCWC)}, 2021, pp. 0987--0992.

\bibitem{ref8}
B.~Perach and S.~Kvatinsky, ``An asynchronous and low-power true random number generator using stt-mtj,'' in \emph{2020 IEEE International Symposium on Circuits and Systems (ISCAS)}, 2020, pp. 1--1.

\bibitem{ref9}
Z.~Hou, M.~Wang, J.~Yin, K.~Shi, B.~Wang, Y.~Zhao, and Z.~Wang, ``High-speed and reconfigurable physical unclonable functions based on sot-mtj array,'' in \emph{2023 IEEE Nanotechnology Materials and Devices Conference (NMDC)}, 2023, pp. 141--144.

\bibitem{ref11}
X.~Jin, W.~Chen, X.~Li, N.~Yin, C.~Wan, M.~Zhao, X.~Han, and Z.~Yu, ``High-reliability, reconfigurable, and fully non-volatile full-adder based on sot-mtj for image processing applications,'' \emph{IEEE Transactions on Circuits and Systems II: Express Briefs}, vol.~70, no.~2, pp. 781--785, 2023.

\bibitem{cnnref1}
M.~Kayed, A.~Anter, and H.~Mohamed, ``Classification of garments from fashion mnist dataset using cnn lenet-5 architecture,'' in \emph{2020 International Conference on Innovative Trends in Communication and Computer Engineering (ITCE)}, 2020, pp. 238--243.

\bibitem{nnref1}
\BIBentryALTinterwordspacing
D.~Zhao, Y.~Zeng, T.~Zhang, M.~Shi, and F.~Zhao, ``Glsnn: A multi-layer spiking neural network based on global feedback alignment and local stdp plasticity,'' \emph{Frontiers in Computational Neuroscience}, vol.~14, 2020. [Online]. Available: \url{https://www.frontiersin.org/articles/10.3389/fncom.2020.576841}
\BIBentrySTDinterwordspacing

\bibitem{nnref2}
X.~Sun, P.~Wang, K.~Ni, S.~Datta, and S.~Yu, ``Exploiting hybrid precision for training and inference: A 2t-1fefet based analog synaptic weight cell,'' in \emph{2018 IEEE International Electron Devices Meeting (IEDM)}, 2018, pp. 3.1.1--3.1.4.

\bibitem{nnref3}
S.-T. Lee, H.~Kim, J.-H. Bae, H.~Yoo, N.~Y. Choi, D.~Kwon, S.~Lim, B.-G. Park, and J.-H. Lee, ``High-density and highly-reliable binary neural networks using nand flash memory cells as synaptic devices,'' in \emph{2019 IEEE International Electron Devices Meeting (IEDM)}, 2019, pp. 38.4.1--38.4.4.

\bibitem{ref13}
Q.~Shao, P.~Li, L.~Liu, H.~Yang, S.~Fukami, A.~Razavi, H.~Wu, K.~Wang, F.~Freimuth, Y.~Mokrousov, M.~D. Stiles, S.~Emori, A.~Hoffmann, J.~Åkerman, K.~Roy, J.-P. Wang, S.-H. Yang, K.~Garello, and W.~Zhang, ``Roadmap of spin–orbit torques,'' \emph{IEEE Transactions on Magnetics}, vol.~57, no.~7, pp. 1--39, 2021.

\end{thebibliography}
\bibliographystyle{IEEEtran}

\vfill

\end{document}